\documentclass[12pt,a4paper]{article}
\usepackage{amsmath,amssymb,amsthm}
\usepackage[margin=1.0in]{geometry}
\usepackage{cite}
\usepackage{graphicx}
\usepackage{enumerate}
\allowdisplaybreaks
\usepackage[colorlinks=true
,urlcolor=blue
,anchorcolor=blue
,citecolor=blue
,filecolor=blue
,linkcolor=blue
,menucolor=blue
,pagecolor=blue
,linktocpage=true
,pdfproducer=medialab
,pdfa=true
]{hyperref}
\numberwithin{equation}{section}

\def\a{\alpha}

\def\g{\gamma}
\def\d{\delta}
\def\e{\epsilon}

\def\k{\kappa}
\def\l{\lambda}
\def\m{\mu}
\def\n{\nu}
\def\x{\xi}
\def\r{\rho}

\def\s{\sigma}
\def\t{\tau}

\def\ph{\phi}

\def\D{\Delta}

\def\Th{\Theta}

\def\S{\Sigma}

\def\L{\Lambda}

\def\be{\begin{equation}}
\def\ee{\end{equation}}
\def\bea{\begin{eqnarray}}
\def\eea{\end{eqnarray}}

\def\pa{\partial}

\def\lp{\left(}
\def\rp{\right)}
\def\ls{\left[}
\def\rs{\right]}
\def\nn{\nonumber}
\def\ie{{\it i.e., }}

\makeatletter
\renewcommand\section{\@startsection {section}{1}{\z@}%
	{-3.5ex \@plus -1ex \@minus -.2ex}
	{2.3ex \@plus.2ex}%
	{\normalfont\large\bfseries}}
\renewcommand\subsection{\@startsection{subsection}{2}{\z@}%
	{-3.25ex\@plus -1ex \@minus -.2ex}%
	{1.5ex \@plus .2ex}%
	{\normalfont\bfseries}}
\makeatother

\begin{document}

\begin{center}
\addtolength{\baselineskip}{.5mm}
\thispagestyle{empty}
\begin{flushright}
\end{flushright}

\vspace{20mm}

{\Large \bf Charged rotating black strings \\ in Einsteinian quartic gravity}
\\[15mm]
{Hamid R. Bakhtiarizadeh\footnote{h.bakhtiarizadeh@kgut.ac.ir}}
\\[5mm]
{\it Department of Nanotechnology, Graduate University of Advanced Technology,\\ Kerman, Iran}

\vspace{20mm}

{\bf  Abstract}
\end{center}

We find an analytic asymptotically anti-de Sitter solution for charged rotating black strings in Einsteinian quartic gravity. By studying the near-horizon behavior of the solutions, we independently extract their thermodynamic properties, analytically. As a consistency check, we show that the first law of thermodynamics for rotating black strings holds exactly in both charged and uncharged cases.

\vfill
\newpage


\section{Introduction}\label{int}

It has been shown that modification of Einstein-Hilbert (EH) action by higher-curvature terms yields a renormalizable theory \cite{Stelle:1976gc}. In the context of string theory, these corrections appear in the $ \a' $ expansion of the low-energy effective actions, which belong to the Lovelock class \cite{Lovelock:1971yv}. A generalization of this class, known as quasi topological gravity (QTG) \cite{Myers:2010ru,Oliva:2010eb,Cisterna:2017umf}, which is also trivial in four dimensions, as Lovelock theory. Recently, a new class of higher-curvature theories known as generalized quasi topological gravity (GQTG) \cite{Hennigar:2017ego} have been introduced, which are neither topological nor trivial in four dimensions, and have the same graviton spectrum on constant curvature backgrounds as general relativity, and therefore, have not negative energy excitations (or ghosts). The equations of motion of these theories for static and spherically symmetric black hole solutions are sufficiently simple to allow a nonperturbative study. The first such theory to be explored was cubic in curvature, and is known as Einsteinian cubic gravity \cite{Bueno:2016xff}. The static and spherically symmetric black hole solutions of this theory are examined in four-dimensional spacetime in \cite{Bueno:2016lrh}. Theories including quartic terms in curvature were introduced shortly afterward \cite{Ahmed:2017jod}, and it was also shown that GQTG theories can be constructed from arbitrarily powers of the curvature, as well as in general dimensions \cite{Bueno:2019ycr}. In \cite{Mir:2019rik}, the authors investigate the thermodynamic behavior of asymptotically anti-de Sitter (AdS) black holes in GQTG containing terms both cubic and quartic in the curvature. As a basic phenomenological test, gravitational lensing effects of spherically symmetric black holes in Einsteinian quartic gravity (EQG) are examined in \cite{Khodabakhshi:2020hny}. Also in \cite{Khodabakhshi:2020ddv}, using a continued fraction ansatz, the authors obtain an analytic approximation for a spherically symmetric black hole solution to EQG. 

The aim of the present paper is to investigate the asymptotically AdS charged rotating black string solutions \cite{Lemos:1994xp,Lemos:1995cm} of quartic version of GQTG. The charged rotating black string solutions have been also studied in \cite{Cai:1996eg,Cardoso:2001vs}, and their thermodynamic properties are investigated in \cite{Dehghani:2002rr,Dehghani:2002jh}. In \cite{Dehghani:2006xt}, a class of charged rotating solutions in $ (n+1) $-dimensional Maxwell-Brans-Dicke theory with flat horizon in the presence of a quadratic potential are constructed and their properties are investigated. A new class of charged rotating dilaton black string solutions in the background of AdS spaces with an appropriate combination of three Liouville-type dilaton potentials are analyzed in \cite{Sheykhi:2008rk}, and their thermodynamics is discussed by using the counterterm method. Also, in \cite{Hendi:2010kv} rotating black string solutions in the presence of two kinds of nonlinear electromagnetic sources, so called Born-Infeld and power Maxwell invariant are investigated and it has been shown that the thermodynamic quantities satisfy the first law of thermodynamics. A new solution for rotating black strings coupled to a nonlinear electromagnetic field in the background of an AdS space is also explored in \cite{Hendi:2013mka}, and the validity of the first law of thermodynamics for solutions is verified.

In EQG theory, there are six quartic curvature combinations, that are nontrivial in four dimensions, and lead to the introduction of six new coupling constants. Here also, as spherical symmetry \cite{Ahmed:2017jod}, imposition of cylindrical symmetry yields a degeneracy insofar as their field equations differ by terms that vanish for a static and cylindrically symmetric metric. In other words, all quartic Lagrangian densities have only a single independent field equation. The solutions are indeed generalization of asymptotically AdS charged rotating black string solutions in four-dimensional ECG  \cite{Bakhtiarizadeh:2021vdo} to include quartic terms in curvature. Note that, although the construction of actions that include $ n $th order terms in curvature from lower order ones via recursive relations has been obtained that allows for construction of any GQTG \cite{Bueno:2019ycr}, but EQG is the highest degree of curvature theories for which an analytic solution of the near-horizon equations for the Hawking temperature and total mass in terms of the horizon radius is possible \cite{Khodabakhshi:2020hny,Khodabakhshi:2020ddv}.

The structure of paper is organized as follows. In Sec. \ref{unch}, we examine the uncharged asymptotically and near a black string horizon solutions in the presence of EQG. By studying the near-horizon behavior, we then obtain exact expressions for the mass and Hawking temperature. We go on to present the contributions to the Wald entropy arising from each of the interactions presented. Finally, we will end this section by computing the angular momentum of black strings. In Sec. \ref{ch}, we present the black string solutions in the context of EQG and in the presence of a Maxwell field. Sec. \ref{dis} is also devoted to discussion.

\section{Asymptotically AdS uncharged solutions}\label{unch}

The action of four-dimensional EQG in the presence of cosmological constant is given by\footnote{Throughout this paper we will work in units of $ G=1 $, so one has to bear in mind that every quantity is expressed in Planck units.}\cite{Ahmed:2017jod}
\be
S=\frac{1}{16\pi}\int d^4x \sqrt{-g}\ls R-2\L-\sum_{i=1}^{6}{\hat \l}_{(i)} S_{4}^{(i)}\rs,\label{action}
\ee
where $ R $ is the usual Ricci scalar, $ \L=-3/{l^2} $ is the negative cosmological constant of AdS space\footnote{The asymptotically de-Sitter solutions can be obtained by taking $ l\rightarrow il $ \cite{Awad:2002cz}.}, and $ {\hat \l}_{(i)} $s are the coupling constants of EQG Lagrangian densities. Quartic-in-curvature corrections to the EH action, $ S_{4}^{(i)} $, are called quasi topological Lagrangian densities, whose analytical expressions read,
\bea
S_{4}^{(1)} &=& - \frac{109}{45} R_{a}{}^{c} R^{ab} R_{b}{}^{d} R_{cd} + \frac{16}{5} R_{ab} R^{ab} R_{cd} R^{cd} + \frac{28}{15} R_{a}{}^{c} R^{ab} R_{bc} R \nonumber \\ 
&& -  \frac{37}{45} R_{ab} R^{ab} \
R^2 + \frac{86}{45} R^{ab} R^{cd} R R_{acbd}  + \frac{1}{18} R^2 R_{abcd} R^{abcd} \nonumber \\ 
&&- \
\frac{208}{15} R_{a}{}^{c} R^{ab} R^{de} \
R_{bdce} + \frac{82}{15} R^{ab} R^{cd} R_{ac}{}^{ef} R_{bdef} -  \frac{7}{30} R R_{ab}{}^{ef} R^{abcd} R_{cdef} \nonumber \\ 
&& -  \frac{8}{9} R_{ab} R^{ab} R_{cdef} \
R^{cdef} -  \frac{1}{3} R^{ab} R_{a}{}^{cde} R_{bc}{}^{fh} R_{defh} + R_{a}{}^{e}{}_{c}{}^{f} R^{abcd} R_{b}{}^{h}{}_{e}{}^{j} R_{dhfj},
\eea
\bea
S_{4}^{(2)} &=& 2 R_{a}{}^{c} R^{ab} R_{b}{}^{d} R_{cd} + \frac{5}{2} R_{ab} R^{ab} R_{cd} \
R^{cd} -  R_{a}{}^{c} R^{ab} R_{bc} R \nonumber \\ 
&& -  \frac{3}{4} R_{ab} R^{ab} R^2 + 4 \
R^{ab} R^{cd} R R_{acbd} + \frac{1}{8} R^2 R_{abcd} R^{abcd} \nonumber \
\\ 
&& - 14 R_{a}{}^{c} R^{ab} R^{de} R_{bdce} + 5 R^{ab} R^{cd} R_{ac}{}^{ef} \
R_{bdef} -  \frac{1}{4} R R_{ab}{}^{ef} \
R^{abcd} R_{cdef} \nonumber \\ 
&& -  \frac{3}{4} R_{ab} R^{ab} R_{cdef} \
R^{cdef} -  R^{ab} R_{a}{}^{cde} R_{bc}{}^{fh} R_{defh} + R_{a}{}^{e}{}_{c}{}^{f} R^{abcd} R_{b}{}^{h}{}_{d}{}^{j} R_{ehfj},
\eea
\bea
S_{4}^{(3)} &=& - \frac{12}{5} R_{a}{}^{c} R^{ab} R_{b}{}^{d} R_{cd} + \frac{2}{5} R_{ab} R^{ab} R_{cd} R^{cd} + \frac{12}{5} R_{a}{}^{c} R^{ab} R_{bc} R \nonumber \\ 
&&-  \frac{1}{5} R_{ab} R^{ab} R^2 + \frac{8}{5} R^{ab} R^{cd} R R_{acbd}  -  \frac{1}{2} R^2 R_{abcd} R^{abcd} \nonumber \\ 
&&- \
\frac{72}{5} R_{a}{}^{c} R^{ab} R^{de} \
R_{bdce} + \frac{48}{5} R^{ab} R^{cd} R_{ac}{}^{ef} R_{bdef}  + \frac{1}{5} R R_{ab}{}^{ef} R^{abcd} \
R_{cdef} \nonumber \\ 
&&+ R_{ab} R^{ab} R_{cdef} \
R^{cdef} - 4 R^{ab} R_{a}{}^{cde} R_{bc}{}^{fh} R_{defh} + R_{ab}{}^{ef} R^{abcd} R_{ce}{}^{hj} \
R_{dfhj},
\eea
\bea
S_{4}^{(4)} &=& - \frac{24}{5} R_{a}{}^{c} R^{ab} R_{b}{}^{d} R_{cd} + \frac{4}{5} R_{ab} R^{ab} R_{cd} R^{cd} + \frac{24}{5} R_{a}{}^{c} R^{ab} R_{bc} R \nonumber \\ 
&&-  \frac{2}{5} R_{ab} R^{ab} R^2 + \frac{16}{5} R^{ab} R^{cd} R \
R_{acbd}  -  R^2 R_{abcd} R^{abcd} \nonumber \\ 
&&-  \frac{144}{5} R_{a}{}^{c} R^{ab} R^{de} R_{bdce} + \frac{96}{5} R^{ab} R^{cd} R_{ac}{}^{ef} R_{bdef} + \frac{2}{5} R R_{ab}{}^{ef} R^{abcd} \
R_{cdef} \nonumber \\ 
&& + 2 R_{ab} R^{ab} R_{cdef} R^{cdef} - 8 R^{ab} R_{a}{}^{cde} \
R_{bc}{}^{fh} R_{defh} + R_{ab}{}^{ef} R^{abcd} R_{cd}{}^{hj} \
R_{efhj},
\eea
\bea
S_{4}^{(5)} &=& - \frac{14}{5} R_{ab} R^{ab} R_{cd} R^{cd} -  \frac{20}{3} R_{a}{}^{b} R_{b}{}^{d} R_{d}{}^{e} R_{e}{}^{a} -  \frac{8}{5} R^{ad} R^{be} R R_{abde} \nonumber \\ 
&&+ \frac{104}{5} R^{ab} R_{c}{}^{e} \
R^{cd} R_{adbe} + R_{ch} R^{ch} R_{abde} R^{abde} + \frac{1}{5} R^2 R_{abde} R^{abde} \nonumber \\ 
&&-  \frac{56}{15} R^{ab} R_{chib} R_{de}{}^{i}{}_{a} R^{dech}  + R_{abd}{}^{c} R^{abde} R_{hife} R^{hif}{}_{c},
\eea
\bea
S_{4}^{(6)} &=& - \frac{308}{15} R_{ab} R^{ab} R_{cd} R^{cd} -  \frac{64}{3} R_{a}{}^{b} R_{b}{}^{d} R_{d}{}^{e} R_{e}{}^{a} + \frac{64}{15} R^{ad} R^{be} R R_{abde} \nonumber \\ 
&&+ \frac{1088}{15} R^{ab} R_{c}{}^{e} \
R^{cd} R_{adbe} + \frac{28}{3} R_{ch} R^{ch} R_{abde} \
R^{abde} -  \frac{8}{15} R^2 R_{abde} R^{abde} \nonumber \\ 
&&-  \frac{224}{15} R^{ab} R_{chib} R_{de}{}^{i}{}_{a} R^{dech} + R_{abcd} R^{abcd} R_{efgh} R^{efgh}.
\eea
We restrict ourselves to the following ansatz \cite{Lemos:1994xp,Lemos:1995cm},
\bea
ds^2=-f(r)g^2(r)\lp \Xi dt -a d\phi \rp ^2+\frac{1}{f(r)}dr^2+\frac{r^2}{l^4} \lp a dt -\Xi l^2 d\phi \rp ^2+\frac{r^2}{l^2}dz^2,\label{met}
\eea
where
\be
\Xi=\sqrt{1+a^2/l^2}.\label{sileq}
\ee
Here the constants $ a $ and $ l $ have dimensions of length and can be regarded as the rotation parameter and the AdS radius, respectively. The ranges of the time and radial coordinates are $ -\infty<t<\infty $, and $ 0\leq r<\infty $. Here, we are going to explore the solutions of Einstein-Maxwell equations with cylindrical symmetry in the presence of Einsteinian quartic gravity. By this we mean spacetimes admitting a commutative two dimensional Lie group $ G_2 $ of isometries. The topology of the two dimensional space, $ t=const. $ and $ r=const. $, generated by $ G_2 $ can be: \begin{enumerate}[(i)] \item the flat torus $ T^2 $ with topology $ S^1\times S^1 $ [\ie $ G_2 =U(1)\times U(1)$] and the ranges $ 0\leq \phi<2\pi,0\leq z<2\pi l $, which describes a closed black string, \item the cylinder with topology $ \mathbb{R}\times S^1 $ [\ie $ G_2 =\mathbb{R}\times U(1)$] and the ranges $ 0\leq \phi<2\pi,-\infty<z<\infty $, which describes a stationary black string, \item the infinite plane with topology $ \mathbb{R}^2 $ and the ranges $ -\infty<\phi<\infty,-\infty<z<\infty $.\end{enumerate} In this paper we consider the topology (ii).\footnote{For a toroidal horizon, the entropy, mass, angular momentum, and charge of the string are obtained from their respective densities by multiplying them by $ 2\pi l $ \cite{Dehghani:2002rr}.} Note that the metric (\ref{met}) is obtained after a boost along the time and angular coordinates from the static black brane solution. Here, the boost is locally defined and the global structure of the metric is different, but this explains why the metric takes the above particular form and why it is not necessary to solve other components of the equations of motion. To find the field equations of action (\ref{action}) on the metric (\ref{met}), we use a method introduced in \cite{Hennigar:2016gkm,Bueno:2016lrh} for static and spherically symmetric spacetimes. By considering the action as a functional of $ f $ and $ g $, \ie $ S[f,g] $, we observe that varying the action with respect to $ f $ yields,
\be
g'(r)=0,
\ee
with the solution $ g = 1 $.\footnote{Without loss of generality we can set $ g(r) = 1 $, for simplicity. In general, one can choose $ g = 1/\sqrt{f_{\infty}} $, where $ f_{\infty}=\lim_{r\rightarrow\infty} f(r) $, to normalize the speed of light on the boundary or in the dual CFT to be $ c = 1 $. However, we can set $ g = 1 $ by reparametrization of time ($ t $) and angular ($ \phi $) coordinates of the metric.} As a result, the EQG theory admits solutions characterized by a single function $ f(r) $. The equation $ \d_{g}S = 0 $ yields, after integrating once, the following equation for $ f $:
\be
-\frac{1}{3}\L r^3-r f-\frac{1}{10}K \ls  \frac{3}{r}f'^4+\frac{4}{r^2}ff'^2\lp f'-3r f'' \rp-\frac{24}{r^3} f^2 f' \lp f'-r f''\rp\rs=r_0,\label{uncheom}
\ee
where $ r_0 $ is an integration constant which is related to the mass of black string as $ r_0= M $. Here, a prime denotes a differentiation with respect to $ r $. Imposition of cylindrical symmetry yields a degeneracy among the different theories in (\ref{action}), in that the constant $ K $ is a linear combination of the six EQG coupling constants,
\be
K\equiv -\frac{5}{6}\sum_{i=1}^{6} \l_{(i)},\label{EQGcc}
\ee
where
\bea
\l_{(1)}&=&-\frac{6}{5}{\hat \l}_{(1)},\quad \l_{(2)}=-3{\hat \l}_{(2)},\quad\l_{(3)}=-\frac{12}{5}{\hat \l}_{(3)},\nn\\ \l_{(4)}&=&-\frac{24}{5}{\hat \l}_{(4)},\quad\l_{(5)}=-\frac{24}{5}{\hat \l}_{(5)},\quad\l_{(6)}=-\frac{96}{5}{\hat \l}_{(6)}.
\eea
The combination (\ref{EQGcc}) appears in (\ref{uncheom}), since each term $ S_4^{(i)} $ has the same contribution to the field equation. This degeneracy shows that there is only one parameter from EQG that can be constrained empirically from analysis of this class of solutions.

\subsection{Large-r asymptotic region}\label{unchasymp}

By setting $ K=0 $ in Eq. (\ref{uncheom}), one finds
\be
f(r)=\frac{r^2}{l^2}-\frac{M}{r},
\ee
where we have set $ \L=-3/{l^2} $ and $ r_0=M $. This is the usual AdS uncharged solution of rotating black string \cite{Lemos:1995cm,Lemos:1994xp}. When $ K\neq 0 $, the asymptotic quantities are modified and to see the changes, we examine the large-$ r $ behavior of the solution. In doing so, we consider the metric function $ f $ as a particular solution in the form of a $ 1/r $ expansion, plus the general solution of the corresponding homogeneous equation, \ie
\be
f=f_{1/r}+f_{h}\quad \text{with} \quad f_{1/r}(r)=\frac{r^2}{l_{\rm eff}^2}+\sum_{n=1}^{\infty}\frac{b_{n}}{r^n}.\label{serexp}
\ee
Substituting this series expansion into Eq. (\ref{uncheom}), one finds the large-$ r $ expansion reads
\be
f_{1/r}(r)=\frac{r^2}{l_{\rm eff}^2}-\frac{M}{\lp 1-\frac{32 K}{5 l_{\rm eff}^6}\rp r}+{\cal O}\lp r^{-3}\rp,\label{f1r}
\ee
where the effective radius of the AdS space, $ {l_{\rm eff}} $, is a solution of the following equation
\be
\frac{8K}{5l_{\rm eff}^8}-\frac{1}{l_{\rm eff}^2}+\frac{1}{l^2}=0.\label{leffeq}
\ee
At the large-$ r $ limit, the linearized homogeneous equation satisfied by $ f_{h}(r) $ is given by 
\be
f_{h}''(r)-\frac{4}{r}f_{h}'(r)-\g^2 r f_{h}(r)=0,\label{uncheq}
\ee
where
\be
\g^2=\frac{\lp 5 l_{\rm eff}^6-32K\rp^2}{180 M Kl_{\rm eff}^8}.\label{gvalue}
\ee
Here, we have kept only the leading terms in the large-$ r $ limit. The solution of Eq. (\ref{uncheq}) in the case of $ \g^2>0 $, is
\be
f_{h}^{(+)}(r)= r^{5/2}\ls A I_{5/3}\lp \frac{2\g r^{3/2}}{3}\rp+B K_{5/3}\lp \frac{2\g r^{3/2}}{3}\rp \rs,
\ee
where $ I $ and $ K $ are the modified Bessel functions of the first and second kinds, and $ A $ and $ B $ are some constants. In the limit of large $ r $ we can approximate the solution by
\be
f_{h}^{(+)}(r)\sim r^{5/2}\ls A \exp\lp \frac{2\g r^{3/2}}{3}\rp+B \exp\lp- \frac{2\g r^{3/2}}{3}\rp \rs,\label{apsol}
\ee 
and so we must set $ A = 0 $ to ensure the AdS boundary conditions are satisfied. As a result, no ghost excitations can propagate to infinity. We shall see shortly that the contribution of the second term can be dismissed. If $ \g^2 < 0 $, then the homogeneous solution asymptotically takes the following form 
\be
f_{h}^{(-)}(r)= r^{5/2}\ls C J_{5/3}\lp \frac{2\left|\g\right| r^{3/2}}{3}\rp+D Y_{5/3}\lp \frac{2\left|\g\right| r^{3/2}}{3}\rp \rs,
\ee
where $ J $ and $ Y $ are the Bessel functions of the first and second kind, and $ C $ and $ D $ are arbitrary constants. In this situation, the solution oscillates rapidly and its amplitude becomes larger than $ r^2/l_{\rm eff}^2 $ at large $ r $. It therefore does not approach AdS asymptotically, and so we must set $ C = D = 0 $ to get rid of this homogeneous part of the solution. For the rest of our considerations, to avoid any oscillating behavior near infinity, we restrict the solutions to the constraint $ \g^2 > 0 $. Finally, we note that the particular solution (\ref{f1r}) polynomially decreases with $ 1/r $ and is the dominant part of the total solution $ f(r) =f_{1/r} + f_h^{(+)} $ for sufficiently large $ r $; we therefore neglect the term $ f_h^{(+)} $ in Eq. (\ref{apsol}) in the sequel.

It is easy to show that the Ricci scalar and the Kretschmann invariant of the spacetime are given by 
\bea
R&=&-f''(r)-4 \frac{f'(r)}{r} -2 \frac{f(r)}{r^2},\nn\\
R_{abcd}R^{abcd}&=&{f''}^{2}(r)+\lp\frac{2f'(r)}{r}\rp^2+\lp\frac{2f(r)}{r^2}\rp^2.
\eea
Since other curvature invariants are a function of $ f''(r) $, $ f'(r)/r $ and $ f(r)/r^2 $, so it is sufficient to study the Ricci and Kretschmann scalars for investigation of the spacetime curvature. Substituting the metric function (\ref{f1r}), one can easily check that for $ r \rightarrow \infty $ the Ricci and Kretschmann scalars go to the values $ R=-12/l_{\rm eff}^2=4\L_{\rm eff} $ and $ R_{abcd}R^{abcd}=24/l_{\rm eff}^4=8/3\L_{\rm eff}^2 $, which by supposing $ {\rm sgn}(\L_{\rm eff}) = {\rm sgn}(\L) $, confirms that our solutions are asymptotically AdS. 

To prove the assumption $ {\rm sgn}(\L_{\rm eff}) = {\rm sgn}(\L) $, let us discuss about the restrictions imposed by Eq. (\ref{leffeq}). If we rewrite this equation in terms of cosmological constant, it takes the following form
\be
\frac{8}{405} K \L_{\rm eff}^4+\frac{\L_{\rm eff}}{3}-\frac{\L}{3}=0.
\ee
Taking the discriminant of the first equation we find that,
\be
\D=-\frac{64 K^2}{1793613375}\lp 3645+2048 K \L^3\rp.
\ee
The discriminant can be either positive, zero, or negative depending on the values of $ K $ and $ \L $:
\bea
\D&>&0\quad {\rm if} \quad \lp \L <0 , K >-\frac{3645}{2048 \L ^3} \rp \quad {\rm or} \quad \lp \L >0 , K <-\frac{3645}{2048 \L ^3} \rp,\nn\\
\D&=&0\quad {\rm if} \quad K=0 \quad {\rm or} \quad K=-\frac{3645}{2048 \L^3},\nn\\
\D&<&0\quad {\rm if} \quad \lp\L <0 ,\lp K <0 \quad {\rm or}\quad 0<K <-\frac{3645}{2048 \L ^3}\rp\rp\nn\\&& \quad~ {\rm or} \quad (\L =0,(K <0\quad {\rm or} \quad K >0))\nn\\&&\quad~ {\rm or} \quad \lp\L >0, \lp-\frac{3645}{2048 \L ^3}<K <0\quad {\rm or} \quad K >0\rp\rp.
\eea
In general, we find that for any given $ \L $ and $ K $, there will only be a single branch that is ghost free. It is also interesting to note that the ghost-free branch has a smooth limit to the Einstein case as $ K\rightarrow 0 $; in other words, the Einstein branch is ghost free. Furthermore, the ghost-free branch has the property that $ {\rm sgn}(\L_{\rm eff}) = {\rm sgn}(\L) $, meaning, for example, if $ \L < 0 $ then the ghost-free branch will possess AdS asymptotics. 

On the other side, in the limit $ r \rightarrow 0 $, both Ricci and Kretschmann scalars diverge, and are finite at $ r\neq 0 $. Therefore, there is a curvature singularity at $ r = 0 $.

\subsection{Near-horizon region}\label{unchnear}

We begin this section by solving the field equation (\ref{uncheom}) near the horizon via a Taylor series expansion.  The general definition of surface gravity is
\be
\k_{g}=\sqrt{-\frac{1}{2}(\nabla^a \chi^b)(\nabla_a \chi_b)}.
\ee
Here, the null generator of the black string horizon is given by $ \chi=\pa_t+\Omega\pa_\phi $, where the angular velocity of the event horizon is given by
\bea 
\Omega=\frac{a}{l^2\Xi}.\label{angve} 
\eea
A simple calculation for the spacetime (\ref{met}) with $ g(r)=1 $ gives $ \k_g =f'(r_h)/2\Xi $. Substituting the ansatz
\be
f(r)=\sum_{n=1}^{\infty} a_n (r-r_h)^n,\label{tay}
\ee
where $ a_n=f^{(n)}(r_h)/n! $, in Eq. (\ref{uncheom}), one finds the two lowest-order equations are given by:
\be
\frac{r_h^3}{l^2}-\frac{24 K \k_g^4 \Xi^4}{5 r_h}-M=0,\label{masseq}
\ee
\be
\frac{3 r_h^2}{l^2}-2 \k_g \Xi r_h-\frac{8 K \k_g^4 \Xi^4}{5 r_h^{2}}=0.\label{surgraeq}
\ee
To solve the above second equation for surface gravity, we follow the approach introduced in \cite{Ahmed:2017jod} for black branes. By substituting $ r_{h} $ from the following definition 
\be
r_{h}\equiv\g_{K}\frac{2 l^2 \Xi}{3} k_{g},
\ee
into Eq. (\ref{surgraeq}), we find that $ \g_{K} $ solves the equation, 
\be
\g_{K}^{4}-\g_{K}^{3}-\frac{27 K}{10 l^6}=0.
\ee 
By studying the above polynomial, we can conclude that there will be real, positive solutions for $ \g_{K} $ provided the coupling satisfies,
\be
-\frac{5 l^6}{128}\leq K \leq 0,\label{constr}
\ee
or
\be
K \geq -\frac{5 l^6}{128},
\ee
with $ K = -5 l^6/128 $ corresponding to a positive, real double root. Note that, only constraint (\ref{constr}) leads to a single root for which Eq. (\ref{surgraeq}) will have a smooth limit to the vacuum of Einstein gravity upon sending coupling constant to zero. Converting $ K\rightarrow-K $ due to the constraint (\ref{constr}) and solving the equations (\ref{masseq}) and (\ref{surgraeq}), one can obtain the surface gravity and mass of black strings, in terms of horizon radius $ r_h $, as
\be
\k_g=\frac{\sqrt[6]{5} r_h \left(\tau ^{3/4}-\sqrt{2 \sqrt{10\xi} l ^{3/2}-\tau ^{3/2}}\right)}{2 \ 2^{5/6} \sqrt[3]{K } \sqrt[6]{\xi }  \sqrt[4]{\tau } \sqrt{l }\Xi},\label{surgra}
\ee
\be
M=\frac{r_h^3}{l ^2} \ls 1+\frac{3 \left(\tau ^{3/4}-\sqrt{2 \sqrt{10\xi} l ^{3/2}-\tau ^{3/2}}\right)^4}{16 \sqrt[3]{10K} \xi ^{2/3} \tau }\rs,\label{mass}
\ee
where 
\be
\x \equiv \sqrt{25 l ^6-640 K}+5 l ^3,\label{kesi}
\ee
\be
\t \equiv \xi ^{2/3}+4 \sqrt[3]{10 K}.\label{tav}
\ee
All remaining $ a_n $, for $ n > 2 $, are determined in terms of $ K $, $ \k_g $, $ r_h $, and $ a_2 $, for which $ a_2 $ is a free parameter.

Note that, in the above quantities for surface gravity and mass, when $ K $ tends to zero and after resolving the ambiguity in the limits via L'Hospital rule, we recover $ k_{g}=3 r_h/2 l^2 \Xi $ and $ M=r_h^3/l^2 $ which are respectively the surface gravity and mass of black strings in Einstein gravity.

\subsection{Thermodynamics}\label{unchthermo}

When higher-curvature corrections are included, the Bekenstein-Hawking entropy is modified by additional terms, which originate from the Wald entropy formula \cite{Wald:1993nt,Iyer:1994ys}
\be
{\sf S}=-2\pi \int_{H} d^2x\sqrt{\g}\frac{\d {\cal L}}{\d R_{abcd}}\e_{ab}\e_{cd},\label{wald}
\ee 
where $ \frac{\d {\cal L}}{\d R_{abcd}} $ is the Euler-Lagrange derivative of gravitational Lagrangian, $ \g $ is the determinant of the induced metric on the horizon, and $ \e_{ab} $ is the binormal of the horizon, normalized by $ \e_{ab} \e^{ab}=-2 $. This antisymmetric tensor is defined by the following relation:
\be
\triangledown_a \chi_b=\k_g\e_{ab}.
\ee
 Using the above equation, the nonzero components of the binormal of horizon of metric (\ref{met}) are given by
\bea
\e_{tr}=-\e_{rt}=-\Xi,\nn\\\e_{r\phi}=-\e_{\phi r}=-a.
\eea   
For the metric (\ref{met}) with a cylindrical horizon placed at $ r = r_h $, one finds the following value for entropy per unit length:
\bea
{\cal S}=\frac{\pi  \Xi  r_h^2 }{10 l }\left[5+\frac{\sqrt{\frac{5}{2}} \left(\tau ^{3/4}-\sqrt{2 \sqrt{10\xi} l ^{3/2}-\tau ^{3/2}}\right)^3}{\sqrt{\xi } l^{3/2} \tau ^{3/4}}\right].\nn\\\label{unchen}
\eea
In writing the above result in terms of horizon radius $ r_h $, the transformation $ K\rightarrow-K $, the relation $ f'(r_h) =2\Xi \k_g $, and Eq. (\ref{surgra}) has been used. Here, in the limit $ K\rightarrow 0 $, we obtain $ {\cal S}=\pi r_h^2 \Xi/2 l $ which is the entropy of black strings in Einstein gravity. The Hawking temperature can easily be found in terms of the horizon radius from surface gravity as
\be
T=\frac{\k_g}{2\pi}=\frac{\sqrt[6]{5} r_h \left(\tau ^{3/4}-\sqrt{2 \sqrt{10\xi} l ^{3/2}-\tau ^{3/2}}\right)}{4 \pi\ 2^{5/6} \sqrt[3]{K } \sqrt[6]{\xi }  \sqrt[4]{\tau } \sqrt{l }\Xi}.\label{temp}
\ee
Here, we are going to find the mass and angular momentum of black strings by using the counterterm method. To this end, we add the Gibbons–Hawking boundary term, which removes the divergences of the action (\ref{action}), as
\be
S_b=\frac{C(l_{\rm eff}^2)}{8\pi}\int_{\pa \cal{M}} d^3x \sqrt{-\g} \Th,
\ee
where $ \g $ is the determinant of the induced metric on the boundary, and $ \Th $ is the trace of the extrinsic curvature $ \Th_{ab} $ of the boundary. Here, $ C(l_{\rm eff}^2) $ is a constant, which depends on the background curvature, and is given by \cite{Bueno:2016ypa}
\bea
C(l_{\rm eff}^2)=-\frac{l_{\rm eff}^2}{6}{\cal L}\vert_{\rm AdS},
\eea
where $ {\cal L}\vert_{\rm AdS} $ is the Lagrangian of the corresponding theory evaluated on $ {\rm AdS}_4 $ background with curvature scale $ l_{\rm eff} $.
We use the counterterm approach \cite{Henningson:1998gx,Balasubramanian:1999re} to eliminate the divergences of action. To do so, we add some local surface integrals to the action to make it finite. The counterterms for the case of EQG, which makes the action finite up to four dimensions, are 
\be
S_{ct}=\frac{C(l_{\rm eff}^2)}{8\pi}\int_{\pa \cal{M}} d^3x \sqrt{-\g} \lp \frac{2}{l_{\rm eff}}-\frac{l_{\rm eff}}{2} {\cal R}\rp,\label{5ctaction}
\ee
where $ {\cal R} $ is the Ricci scalar for the boundary metric $ \g $. Notice that, $ l_{\rm eff} $ is a scale length factor that depends on $ l $ and $ K $, that must reduce to $ l $ as $ K $ goes to zero. That is indeed the root of Eq. (\ref{leffeq}). The total action can be written as a linear combination of the action of bulk, boundary and the counterterm,
\be
S_{total}=S+S_{b}+S_{ct}.
\ee 
Having had the total finite action, one can use the Brown-York definition of stress tensor \cite{Brown:1992br} by varying the action with respect to boundary metric $ \g_{ab} $, and finding a divergence-free stress tensor
\be
T^{ab}=\frac{ 1-\frac{32 K}{5 l_{\rm eff}^6}}{8\pi}\lp \Th^{ab}-\Th \g^{ab}+\frac{2}{l_{\rm eff}}\g^{ab}-\frac{l_{\rm eff}}{2} {\cal G}^{ab}\rp,\label{enmomtens}
\ee
where $ {\cal G}_{ab}={\cal R}_{ab}-{\cal R}\g_{ab}/2 $ is the Einstein tensor of boundary metric $ \g_{ab} $. For asymptotically AdS solutions with flat horizons $ {\cal R}_{abcd}(\g) = 0 $, the only nonvanishing counterterm, is the first term in (\ref{5ctaction}), which yields the stress tensor (\ref{enmomtens}) up to the third term. Using (\ref{enmomtens}), one can define the quasilocal conserved quantities for an asymptotically AdS spacetime. The conserved charge associated to a Killing vector $ \xi_a $ is given by
\be
Q_\xi=\int_{\S} d^3x\sqrt{\s} u^a T_{ab}\xi^b.\label{consch}
\ee
Here, $ u_a =-N\d^{0}_{a} $, while $ N $ and $ \s $ are the lapse function and the metric of spacelike surface $ \S $, which appear in the ADM–like decomposition of the boundary metric,
\be
\g_{ij}dx^i dx^j=-N^2 dt^2+\s_{ab}\lp dx^a + V^a dt\rp \lp dx^b + V^b dt \rp,
\ee
where $ V^a $ is the shift vector. To obtain the total mass (energy), we should set $ \xi=\pa_t $, \ie the Killing vector conjugate to time coordinate $ t $, and to obtain the angular momentum, we should set $ \xi=\pa_{\ph} $, \ie the Killing vector conjugate to angular coordinate $ \ph $. Using the definition (\ref{consch}) for conserved charges, we find the total mass per unit length of horizon to be given by
\be
{\cal M}=\frac{1}{8l}\lp 3 \Xi^2-1 \rp M,\label{mas}
\ee
while the angular momentum per unit length of horizon is given by
\be
{\cal J}=\frac{3}{8l} \Xi a M=\frac{3}{8} \Xi\sqrt{\Xi^2-1} M,\label{ang}
\ee
where Eq. (\ref{sileq}) has been used. Substituting the mass (\ref{mass}) into the above equations, one arrives at   
\bea
{\cal M}=\frac{r_h^3}{8l ^3} \lp 3 \Xi^2-1 \rp\ls 1+\frac{3 \left(\tau ^{3/4}-\sqrt{2 \sqrt{10\xi} l ^{3/2}-\tau ^{3/2}}\right)^4}{16 \sqrt[3]{10K} \xi ^{2/3} \tau }\rs,\label{unchmass}
\eea
\bea
{\cal J}=\frac{3r_h^3}{8l ^2} \sqrt{\Xi^2-1}\ls 1+\frac{3 \left(\tau ^{3/4}-\sqrt{2 \sqrt{10\xi} l ^{3/2}-\tau ^{3/2}}\right)^4}{16 \sqrt[3]{10K} \xi ^{2/3} \tau }\rs.\label{unchang}
\eea

To check the first law of thermodynamics for uncharged black strings, we consider $ {\cal M}(r_h) $ in Eq. (\ref{unchmass}) as a Smarr-type formula. On the other side, from Eqs. (\ref{unchen}) and (\ref{unchang}), one can also obtain $ r_h=r_h({\cal S},{\cal J}) $ and as a result $ {\cal M}={\cal M}({\cal S},{\cal J}) $. Therefore, temperature $ T $ and angular velocity $ \Omega $ can be defined as intensive quantities conjugate to $ {\cal S} $ and $ {\cal J} $, respectively, and can be obtained from 
\bea 
T\equiv\lp\frac{\pa {\cal M}}{\pa {\cal S}}\rp_{{\cal J}}&=&\lp\frac{\pa {\cal M}}{\pa r_h}\rp_{{\cal J}}\lp\frac{\pa {\cal S}}{\pa r_h}\rp^{-1}_{{\cal J}},\nn\\
\Omega\equiv\lp\frac{\pa {\cal M}}{\pa {\cal J}}\rp_{{\cal S}}&=&\lp\frac{\pa {\cal M}}{\pa r_h}\rp_{{\cal S}}\lp\frac{\pa {\cal J}}{\pa r_h}\rp^{-1}_{{\cal S}}.
\eea
Our calculations show that $ T $ and $ \Omega $ calculated by the above equations coincide with corresponding ones in (\ref{temp}) and (\ref{angve}), respectively. So, as a consistency check, one can verify that the first law of thermodynamics for uncharged rotating black strings, 
\be
d{\cal M}=Td{\cal S}+\Omega d{\cal J},
\ee
holds exactly.
\section{Asymptotically AdS charged solutions}\label{ch}

Here we add a Maxwell term to the EQG action
\be
S=\int d^4x \sqrt{-g}\ls \frac{1}{16\pi}\lp R-2\L-\sum_{i=1}^{6}{\hat \l}_{(i)} S_{4}^{(i)}\rp-\frac{1}{4}F_{ab}F^{ab}\rs,\label{chaction}
\ee
where $ F_{ab}=2\pa_{[a} A_{b]} $. The vector potential and nonzero components of electromagnetic tensor are \cite{Lemos:1995cm}
\bea
&&A_{a}=A_0(r)\lp \Xi \d_{a}^{t}- a \d_{a}^{\phi} \rp;\nn\\&&F_{tr}=-F_{rt}=-\Xi A'_0(r),F_{\phi r}=-F_{r\phi}=a A'_0(r).\label{nonvanF}
\eea
The $ t $ and $ \phi $ components of the Maxwell equation, $ \triangledown_a F^{ab}=0 $, now read
\be
r A''_0 g-r A'_0 g'+2 A'_0 g =0.
\ee
Here also we find that $ g'=0 $ and accordingly set $ g=1 $, which yields the following value for the gauge potential
\be
A_0(r)=\frac{q}{4 \pi r},
\ee
where $ q $ is an integration constant, which is indeed the electric charge of the black string. As the uncharged case, the equation for $ f $ can be found by variation of (\ref{chaction}) with respect to the metric function $ g $, 
\be
-\frac{1}{3}\L r^3+\frac{Q^2}{r}-r f-\frac{1}{10}K \ls  \frac{3}{r}f'^3+\frac{4}{r^2}ff'^2\lp f'-3r f'' \rp-\frac{24}{r^3} f^2 f' \lp f'-r f''\rp\rs=r_0,\label{cheom}
\ee
where we have defined $ Q^2\equiv q^2/4\pi $, and as the uncharged case $ r_0 $ is an integration constant, which is related to the mass of the black string as $ r_0=M $. 

\subsection{Large-r asymptotic region}\label{chasymp}

In this section, we will perform an asymptotic expansion around $ r\rightarrow \infty $. Setting $ K=0 $ in Eq. (\ref{cheom}) leads to
\be
f(r)=\frac{r^2}{l^2}-\frac{M}{r}+\frac{Q^2}{r^2},
\ee
which is the usual asymptotically AdS charged solution of rotating black strings \cite{Lemos:1995cm,Lemos:1994xp}. To examine the large-$ r $ behavior of the solution, suppose that the metric function $ f $ is given by (\ref{serexp}). By inserting this series expansion into Eq. (\ref{cheom}), one observes that the large-$ r $ expansion of metric function $ f $ takes the following form
\be
f_{1/r}(r)=\frac{r^2}{l_{\rm eff}^2}-\frac{M}{\lp 1-\frac{32 K}{5 l_{\rm eff}^6}\rp r}+\frac{Q^2}{\lp 1-\frac{32 K}{5 l_{\rm eff}^6}\rp r^2}+{\cal O}(r^{-3}),\label{largerexp}
\ee
where the effective radius of the AdS space $ l_{\rm eff} $, is defined by Eq. (\ref{leffeq}). The linearized homogeneous equation satisfied by $ f_{h}(r) $ is given by
\be
f_{h}''(r)+\frac{12 M}{4 Q^2 -3 M r}f_{h}'(r)+\frac{3 M \g^2 r^2}{4 Q^2 -3 M r}f_{h}(r) =0,
\ee
at the large-$ r $ limit. Notice that when $ Q=0 $, we recover Eq. (\ref{uncheq}). Unfortunately, this equation can not be solved analytically. But due to the existence of $ \lp 4 Q^2 -3 M r \rp $ in the denominator, it is possible to expand the above equation at large $ r $ one more time and find a solution and discuss the boundary conditions. By doing so, we arrive at 
\be
f_{h}''(r)- \frac{\lp 4 Q^2+3 M r\rp \g^2}{3 M} f_{h}(r)=0.
\ee
Here, $ \g^2 $ is given by Eq. (\ref{gvalue}). The solution reads,
\be
f_{h}(r)= A' \text{Ai}\left(\frac{4 Q^2+3 M r}{3 M}\left|\gamma^2\right|^{1/3}\right)+B' \text{Bi}\left(\frac{4 Q^2+3 M r}{3 M}\left|\gamma^2\right|^{1/3}\right),
\ee
where $ \text{Ai} $ and $ \text{Bi} $ are the Airy functions of the first and second kinds, and $ A' $ and $ B' $ are some constants. In the case of $ \g^2>0 $, and at large $ r $,
\bea
f_{h}^{(+)}(r)&\sim & \left(\frac{4 Q^2+3 M r}{3 M}\gamma ^{2/3}\right)^{1/4}\times\\&&\left\{ A' \exp\ls-\frac{2}{3}\left(\frac{4 Q^2+3 M r}{3 M}\gamma ^{2/3}\right)^{3/2}\rs+B' \exp\ls \frac{2}{3}\left(\frac{4 Q^2+3 M r}{3 M}\gamma ^{2/3}\right)^{3/2}\rs \right\},\nn\label{apsol1}
\eea
and so we must set $ B' = 0 $ to ensure the AdS boundary conditions are satisfied. As a result no ghost excitations can propagate to infinity. If $ \g^2 < 0 $, then the homogeneous solution asymptotically takes the following form at large $ r $:
\bea
f_{h}^{(-)}(r)&\sim & \left(\frac{4 Q^2+3 M r}{3 M}\gamma ^{2/3}\right)^{-1/4}\times\\&&\left\{ A' \sin\ls\frac{2}{3}\left(\frac{4 Q^2+3 M r}{3 M}\gamma ^{2/3}\right)^{3/2}\rs+B' \cos\ls \frac{2}{3}\left(\frac{4 Q^2+3 M r}{3 M}\gamma ^{2/3}\right)^{3/2}\rs \right\}.\nn
\eea
Here also the solution oscillates rapidly, and therefore, does not approach AdS asymptotically, and so we set $ A' = B' = 0 $. As the uncharged case, to avoid any oscillating behavior at infinity, we restrict the solutions to the constraint $ \g^2 > 0 $. Observe that all the leading asymptotic corrections to the metric come from the solution (\ref{largerexp}), while the contributions from the homogeneous equation are extremely subleading. Hence, the term proportional to $ A' $ above can be discarded from the asymptotic expansion.

\subsection{Near-horizon region}\label{chnear}

Expanding the field equation (\ref{cheom}) near the horizon, via the Taylor series (\ref{tay}), leads to the following equations at two lowest-order:
\be
\frac{r_h^3}{l^2}-\frac{24 K \k_g^4 \Xi^4}{5 r_h}+\frac{Q^2}{r_h}-M=0,
\ee
\be
\frac{3 r_h^2}{l^2}-\frac{8 K \k_g^4 \Xi^4}{5 r_h^{2}}-\frac{Q^2}{r_h^2}-2 \k_g \Xi r_h=0.
\ee 
By converting $ K\rightarrow-K $ because of the constraint (\ref{constr}) and solving the above equations, one can easily find the surface gravity and mass as
\be
\k_g=\frac{\sqrt[6]{5} \left(\sqrt{\frac{6 \sqrt{10\tau '} r_h^3 l^{3/2}}{\sqrt{\eta '}}- \eta '}+ \sqrt{\eta '}\right)}{2 \ 2^{5/6} \sqrt[3]{3K} \Xi  \sqrt{l} \sqrt[6]{\tau '}},\label{chsurgra}
\ee
\be
M=\frac{r_h^3}{l^2}+\frac{Q^2}{r_h}+\frac{\left(\sqrt{\frac{6 \sqrt{10\tau '} r_h^3 l^{3/2}}{\sqrt{\eta '}}- \eta '}+ \sqrt{\eta '}\right)^4}{16 \sqrt[3]{30K} r_h l^2 \tau '^{2/3}},\label{chmass}
\ee
where
\be
\xi'\equiv 2025 r_h^{12} l^6-1920 K \left(3 r_h^4-Q^2 l ^2\right)^3, 
\ee
\be
\tau'\equiv \sqrt{\xi '}+45 r_h^6 l^3,
\ee
\be
\zeta'\equiv 4 \sqrt[3]{30} \left(3 r_h^4-Q^2 l ^2\right).
\ee
\be
\eta'\equiv \sqrt[3]{K} \zeta '+\tau '^{2/3}.
\ee
As mentioned above for the uncharged case, all $ a_n $ for $ n > 2 $ are determined in terms of $ K $, $ Q $, $ r_h $ and $ a_2 $, while $ a_2 $ is left undetermined by the field equations.

As a consistency check, when $ K\rightarrow 0 $ the above quantities for surface gravity and mass tend to $ k_{g}=\frac{3 r_h}{2 l^2\Xi}-\frac{Q^2}{2 r_h^3 \Xi} $ and $ M=\frac{r_h^3}{l^2}+\frac{Q^2}{r_h} $, respectively, which are the corresponding ones in RN-AdS solutions of black strings. 

\subsection{Thermodynamics}\label{chthermo}

In the case of charged solutions, we find that the Wald entropy per unit length reads,
\be
{\cal S}=\frac{\pi  \Xi  \ls 5 r_h^4+\frac{\sqrt{\frac{5}{2}} r_h \left(\sqrt{\frac{6 \sqrt{10\tau '} r_h^3 l^{3/2}}{\sqrt{\eta '}}- \eta '}+ \sqrt{\eta '}\right)^3}{3 l ^{3/2} \sqrt{\tau '}}\rs}{10 r_h^2 l}.\label{chen}
\ee
Here also we observe that when coupling constant goes to zero, we recover the entropy of black strings in Einstein-Maxwell gravity, as expected. The Hawking temperature also can easily be found in terms of total charge and the horizon radius by multiplying the surface gravity (\ref{chsurgra}) with a factor $ 1/2\pi $, 
\be
T=\frac{\sqrt[6]{5} \left(\sqrt{\frac{6 \sqrt{10\tau '} r_h^3 l^{3/2}}{\sqrt{\eta '}}- \eta '}+ \sqrt{\eta '}\right)}{4\pi \ 2^{5/6}  \sqrt[3]{3K} \Xi  \sqrt{l} \sqrt[6]{\tau '}}.\label{chtemp}
\ee
Substituting Eq. (\ref{chmass}) into the first line of Eqs. (\ref{mas}) and (\ref{ang}), one finds the total mass (energy) and angular momentum per unit length as
\be
{\cal M}=\frac{1}{8l}\lp 3 \Xi^2-1 \rp \ls \frac{r_h^3}{l^2}+\frac{Q^2}{r_h}+\frac{\left(\sqrt{\frac{6 \sqrt{10\tau '} r_h^3 l^{3/2}}{\sqrt{\eta '}}- \eta '}+ \sqrt{\eta '}\right)^4}{16 \sqrt[3]{30K} r_h l^2 \tau '^{2/3}}\rs,
\ee
\be
{\cal J}=\frac{3}{8} \sqrt{\Xi^2-1} \ls \frac{r_h^3}{l^2}+\frac{Q^2}{r_h}+\frac{\left(\sqrt{\frac{6 \sqrt{10\tau '} r_h^3 l^{3/2}}{\sqrt{\eta '}}- \eta '}+ \sqrt{\eta '}\right)^4}{16 \sqrt[3]{30K} r_h l^2 \tau '^{2/3}} \rs.
\ee
Now we are going to find the electric potential and total charge of black strings. The electric potential $ \Phi $, measured at infinity with respect to the horizon, is defined by \cite{Dehghani:2002rr}
\be
\Phi=A_a \chi^a\lvert_{r\rightarrow\infty}-A_a\chi^a\lvert_{r=r_h},
\ee
Calculating the above expression yields the following value for electric potential
\bea
\Phi=\frac{q}{4 \pi  \Xi  r_h}=\frac{Q}{2 \sqrt{\pi } \Xi  r_h}.\label{chpot} 
\eea
To calculate the electric charge of the black string, we first determine the electric field by considering the projections of the electromagnetic field tensor on a special hypersurface. The normal vectors to such a hypersurface are
\be
u^0 = \frac{1}{N} ,u^r = 0, u^i=-\frac{V^i}{N},
\ee
where $ N $ and $ V^i $ are the lapse function and shift vector, respectively. The electric field is also given by $ E^\m = g^{\m\r}F_{\r\n}u^{\n} $. The electric charge per unit length of black strings can be found by calculating the flux of the electric field at infinity, yielding
\bea
{\cal Q}=\frac{q \Xi}{2l}=\frac{\sqrt{\pi } Q \Xi}{l}.\label{totcha}
\eea

Now, we are in a position to verify the validity of the first law of thermodynamics for charged black strings. To do so, we first consider Eq. (\ref{chmass}) as a Smarr-type formula for mass $ {\cal M}(r_h,{\cal Q}) $. It can also be seen from Eqs. (\ref{chen}), (\ref{unchang}) and (\ref{totcha}) that $ r_h=r_h({\cal S},{\cal J},{\cal Q}) $ and accordingly $ {\cal M}={\cal M}({\cal S},{\cal J},{\cal Q}) $. Therefore, temperature $ T $, angular velocity $ \Omega $ and electric potential $ \Phi $ can be defined as conjugate intensive parameters for $ {\cal S} $, $ {\cal J} $ and $ {\cal Q} $, and are given by the following expressions
\bea 
T\equiv\lp\frac{\pa {\cal M}}{\pa {\cal S}}\rp_{{\cal J},{\cal Q}}&=&\lp\frac{\pa {\cal M}}{\pa r_h}\rp_{{\cal J},{\cal Q}}\lp\frac{\pa {\cal S}}{\pa r_h}\rp^{-1}_{{\cal J},{\cal Q}},\nn\\ \Omega\equiv\lp\frac{\pa {\cal M}}{\pa {\cal J}}\rp_{{\cal S},{\cal Q}}&=&\lp\frac{\pa {\cal M}}{\pa r_h}\rp_{{\cal S},{\cal Q}}\lp\frac{\pa {\cal J}}{\pa r_h}\rp^{-1}_{{\cal S},{\cal Q}},\nn\\ \Phi\equiv\lp\frac{\pa {\cal M}}{\pa {\cal Q}}\rp_{{\cal S},{\cal J}}&=&\lp\frac{\pa {\cal M}}{\pa {\cal Q}}\rp_{r_h,{\cal J}}+\lp\frac{\pa {\cal M}}{\pa r_h}\rp_{{\cal Q},{\cal J}}\lp\frac{\pa r_h}{\pa {\cal Q}}\rp_{{\cal S},{\cal J}}\nn\\&=&\lp\frac{\pa {\cal M}}{\pa {\cal Q}}\rp_{r_h,{\cal J}}-\frac{\lp\frac{\pa {\cal S}}{\pa {\cal Q}}\rp_{r_h,{\cal J}}\lp\frac{\pa {\cal M}}{\pa r_h}\rp_{{\cal Q},{\cal J}}}{\lp\frac{\pa {\cal S}}{\pa r_h}\rp_{{\cal Q},{\cal J}}}\nn\\&=&\lp\frac{\pa {\cal M}}{\pa {\cal Q}}\rp_{r_h,{\cal J}}-T \lp\frac{\pa {\cal S}}{\pa {\cal Q}}\rp_{r_h,{\cal J}}.
\eea
It is not difficult to show that the above quantities coincide with those obtained in Eqs. (\ref{chtemp}), (\ref{angve}) and (\ref{chpot}). Having had the thermodynamic properties, one can easily check that the first law of thermodynamics for charged rotating black strings, 
\be
d{\cal M}=Td{\cal S}+\Omega d{\cal J}+\Phi d{\cal Q},
\ee
is fully satisfied in the presence of EQG theory.

\section{Discussion}\label{dis}

We have done the first study of asymptotically AdS charged rotating black string solutions in the context of EQG, a class of theories for which corrections to the EH action are quartic in curvature. It can be seen that under cylindrical symmetry, just like spherical one, there is a degeneracy that yields the same field equation for all theories in the class. It means that the corrections to Einstein gravity depend only on a single coupling constant. These theories are the next simplest kinds that occur in GQTG after Einsteinian cubic gravity, and are the highest in curvature that allow explicit solutions for the total mass and Hawking temperature in terms of the horizon radius. We have obtained the solutions in both the near-horizon and large-distance approximations.

We have proved that, in our model, the ghost-free branch of solutions will occur when the spacetime is asymptotically AdS, as well. We have found that the solutions have a curvature singularity in $ r=0 $ at which both Ricci and Kretschmann scalars diverge. We have used the definition of conserved charge provided by the quasilocal stress–tensor introduced by Brown and York to calculate the total energy (mass) and angular momentum of these solutions. We have also obtained exact formulas for all the relevant thermodynamical quantities and using them the validity of first law of thermodynamics for rotating black strings have been verified in both charged and uncharged cases. This is an interesting check of our calculations, since the physical quantities appearing in this expression namely, the Abbott-Deser mass $ {\cal M} $, the Wald entropy $ {\cal S} $, the Hawking temperature $ T $, the angular velocity $ \Omega $, the angular momentum $ {\cal J} $, the electric potential $ \Phi $ and the charge $ {\cal Q} $ have been computed independently. 

As an extra check of our calculations, it also can be seen that when the EQG coupling constant $ K $ tends to zero, we recover the corresponding physical quantities in Einstein gravity. 

For future works, it also would be interesting to perform this kind of study in higher dimensions, \ie finding the black brane solutions in the context of EQG. Finding the solutions in the presence of nonlinear Born-Infeld electrodynamics is also an interesting feature. More generally, it would be of interest to study holographic implications of the quartic curvature term.
\appendix

\section*{Acknowledgements}\addcontentsline{toc}{section}{Acknowledgements}

We would like to thank the authors of Ref. \cite{Nutma:2013zea} for developing the excellent \emph{Mathematica} package ``xTras,'' which we have used extensively for symbolic calculations.


\providecommand{\href}[2]{#2}\begingroup\raggedright
\endgroup
\end{document}